\documentclass[prb,a4paper,showpacs,twocolumn,superscriptaddress,longbibliography]{revtex4-1}

\usepackage{amssymb}
\usepackage{amsmath}
\usepackage{amsfonts}
\usepackage{graphicx}
\usepackage{bm}
\usepackage{color}
\usepackage{multirow}
\usepackage{natbib}
\usepackage{hyperref}

\DeclareMathOperator{\sgn}{sgn}

\DeclareMathOperator{\im}{Im}
\DeclareMathOperator{\re}{Re}

\begin{document}

\title{Two-instanton approximation to the Coulomb blockade problem}

\author{I. S.~Burmistrov}

\affiliation{L.D. Landau Institute for Theoretical Physics, Kosygina
  street 2, 117940 Moscow, Russia}
  
\affiliation{Moscow Institute of Physics and Technology, 141700 Moscow, Russia}

\affiliation{Condensed-matter Physics Laboratory, National Research University Higher School of Economics, 101000 Moscow, Russia}

\begin{abstract}
We develop the two-instanton approximation to the current-voltage characteristic of a single electron transistor within the Ambegaokar-Eckern-Sch\"on model. We determine the temperature and gate voltage dependence of the Coulomb blockade oscillations of the conductance and the effective charge. We find that a small (in comparison with the charging energy) bias voltage  leads to significant suppression of the Coulomb blockade oscillations and to appearance of the bias-dependent phase shift. 
 \end{abstract}


\maketitle

\section{Introduction}
\label{s1}

For several decades Coulomb blockade remains a powerful tool for
observation of interaction and quantum effects in single electron devices [\onlinecite{zaikin,ZPhys,grabert,blanter,aleiner,Glazman}]. In particular, Coulomb blockade restricts 
an electron transport through a single electron transistor (SET) at low temperatures ($T$). It is the Coulomb energy $E_c=e^2/2C$ where $C$ denotes the total capacitance of a SET, 
that is responsible for Coulomb blockade at $T\ll E_c$. By changing the gate voltage $U_g$ one can induced the external charge on the island, $q= C_g U_g/e$, where $C_g$ denotes the gate capacitance. The tunnel junction between the source (drain) electrode and the island of a SET is characterized by the dimensionless (in units $e^2/h$) conductance $g_s$ ($g_d$). For $g_{s,d}\ll 1$ the orthodox theory of Coulomb blockade predicts the maximum of the SET conductance ($G$) at integer values of $q$  [\onlinecite{OTCB}]. In the opposite case $g_{s}\gg 1$ or $g_d\gg 1$, the conductance of a SET has weak Coulomb blockade oscillations with $q$ [\onlinecite{Cond}].  

Coulomb blockade can be conveniently described in the framework of the Ambegaokar-Eckern-Sch\"{o}n (AES) model [\onlinecite{AES}]. In spite of well-known limitations for its application to a realistic SET [\onlinecite{Falci,BEAH,ET,Glazman}], the AES model adequately describes Coulomb blockade in the limits of weak and strong tunneling. The AES model arises also as an effective description of a quantum particle on a ring in the presence of the ohmic dissipative environment [\onlinecite{PR}]. In completely different context the AES model is known as the ``circular brane model'' [\onlinecite{LZ}].

Physically, the AES model describes spatially independent fluctuations of a voltage on the SET island. Due to $U(1)$ nature of the corresponding boson variable in the imaginary time, the AES model possesses a non-trivial topology which admits 
topologically non-trivial solutions of the classical equations of motion (\emph{Korshunov instantons})
[\onlinecite{Instantons1},\onlinecite{Instantons11}]. It was shown that non-trivial topology of the AES model results in existence of the effective charge $\mathcal{Q}$ which is integer quantized in the limit of zero temperature [\onlinecite{Burmistrov1},\onlinecite{Burmistrov2}]. The effective charge is expressed via the average charge on the island and the anti-symmetrized current noise. The SET conductance  and the effective charge  are analogous to the longitudinal and Hall conductances in the theory of the integer quantum Hall effect [\onlinecite{PruiskenBurmistrov2005}]. Recently, the physical meaning of $\mathcal{Q}$ in the AES model has been elucidated for a problem of a quantum particle on a ring in the presence of dissipative environment. It was demonstrated [\onlinecite{Semenov}] that the integer quantization of the effective charge is related to the conservation of the angular moment of the total system: the particle and the environment. 

Present work is motivated by recent experiments on electron transport via 
a gold nanoparticle which are capacitively coupled to the gate electrode and with controllable varying coupling to the source and drain electrodes  [\onlinecite{Frydman2015}]. The dependence of conductance on a gate voltage at a given temperature for a set of such SETs has been measured. The samples were typically characterized by large total tunneling conductance, $g=g_s+g_d \gtrsim 1$. In several cases non-sinusoidal oscillations of conductance with $U_g$ has been observed. In Ref. [\onlinecite{Frydman2015}] these oscillations were attributed to instanton configurations of the voltage fluctuations in the nanoparticle. However, the detailed theory of these non-sinusoidal Coulomb blockade oscillations of the conductance has not been developed so far. 

In this paper, we fill this gap and compute the response function within the two-instanton approximation to the AES model. The imaginary part of the response function determines the current-voltage characteristics and, consequently, the conductance. The real part of the response function together with the average charge determines the effective charge. We demonstrate that the amplitude of harmonics of oscillations of the conductance and the effective charge with $q$  is indeed controlled by a small parameter, $g^2 E_c e^{-g/2}/(2\pi T) \ll 1$, as it was proposed in Ref. [\onlinecite{Cond}] on the basis of one-instanton computations. In particular, this implies that in the case of weak Coulomb blockade oscillations, $g^2 E_c e^{-g/2}/(2\pi T) \ll 1$, the amplitude of the second harmonic is much smaller than the amplitude of the first one. For a small bias voltage, $|eV| \ll E_c$, the amplitude of the Coulomb blockade oscillations in the current-voltage characteristics is controlled by a small parameter  $g^2 E_c e^{-g/2}/\max\{2\pi T, |eV|\} \ll 1$. There is suppression of the Coulomb blockade oscillations with increase of the bias voltage. 

The structure of the paper is as follows. In Sec. \ref{s2} we introduce the AES action and physical observables for description of a SET. The one and two instanton analysis of the response function and physical observables is presented in Sec. \ref{s3}. Discussion of our results and conclusions are presented in Sec. \ref{s4}.  Some additional details are given in Appendix \ref{app1}. We use units with $\hbar=e=1$ through out the paper except for a some final results.

\section{Formalism}
\label{s2}

\subsection{The action}

As well-known [\onlinecite{Falci,BEAH,ET,Glazman}], a SET can be described by the AES action provided the following assumptions are satisfied: (i) the number of channels in a tunnel junction is large, (ii) each channel is in the tunneling regime, (iii) the level spacing for a single-particle states in the island is small compared to  thetemperature, (iv) the Thouless energy is the largest energy scale in the problem. The effective action in the imaginary time is given as~\cite{AES}
\begin{equation}
\label{eq:Zstart}
 Z = \int \mathcal{D}[\phi]\, e^{-\mathcal{S} [\phi]}, \quad \mathcal{S} [\phi] =
 \mathcal{S}_d +\mathcal{S}_t + \mathcal{S}_c.
\end{equation}
Here the part ($\beta=1/T$)
\begin{equation}\label{eq: SdStart}
 \mathcal{S}_d[\phi] = \frac{g}{4}\int\limits_{0}^{\beta} d\tau_1 d\tau_2\,
 \alpha(\tau_{1}-\tau_{2})\,e^{-i \phi(\tau_1)+i\phi(\tau_2)}
\end{equation}
takes into account the tunneling of electrons between the
island and the reservoirs. It involves the nonlocal in imaginary time kernel 
\begin{equation}
\alpha(\tau)=\frac{T}{\pi}\sum_{\omega_n}|\omega_n|e^{-i\omega_n\tau} ,
\end{equation}
where $\omega_n=2\pi T n$. The term
\begin{equation}
 \mathcal{S}_t[\phi] = - i q \int\limits_{0}^{\beta}d\tau
\dot{\phi}.
\end{equation}
describes the capacitive coupling between the island and the gate. The Coulomb interaction between electrons on the island
is taken into account by the term 
\begin{equation}\label{eq:StcStart}
 \mathcal{S}_c[\phi] = \frac{1}{4 E_c} \int_{0}^{\beta} d \tau\, \dot{\phi}^2 .
\end{equation}

\subsection{Physical observables}

The fundamental physical observable for a SET is the conductance $G$. It can be written as
\begin{equation}
G = \frac{e^2}{h} \frac{g_l g_r}{(g_l+g_r)^2} \mathcal{G},
\end{equation}
where 
\begin{equation}
\mathcal{G} = 4\pi \im \frac{\partial K^R(\omega)}{\partial \omega} \Biggl |_{\omega\to 0}.
\label{eq:defG}
\end{equation}
Here the retarded correlation function $K^R(\omega)$ can be obtained from the Matsubara function
\begin{equation}
K(i\omega_n) = \frac{g T}{4}\sum_{\omega_m} \alpha(i\omega_{m+n}) D(i\omega_m) ,
\label{eq:Kdef}
\end{equation}
where
\begin{equation}
D(i\omega_n) = \int\limits_0^\beta d\tau\, e^{i\omega_n\tau} \left \langle e^{-i\phi(\tau_1)+i\phi(0)} \right \rangle  
\label{eq:Ddef}
\end{equation}
stands for the two-point correlation function of the Coulomb boson.

In Refs. [\onlinecite{Burmistrov1},\onlinecite{Burmistrov2}] it was shown that in addition to the conductance a SET is characterized by  
the effective charge
\begin{equation}
\mathcal{Q}= Q + \re \frac{\partial K^R(\omega)}{\partial \omega} \Biggl |_{\omega\to 0}.
\label{eq:defQ}
\end{equation}
We remind that $\mathcal{Q}$ should be contrasted with the average charge on the island, 
\begin{equation}
Q = q - \frac{T}{2E_c} \frac{\partial \ln Z}{\partial q} .
\label{eq:def-avQ}
\end{equation}

\section{Instanton analysis}
\label{s3}

\subsection{Korshunov instantons}

The dissipative part $\mathcal{S}_d$ of the AES action has classical finite action solutions $\phi_W (\tau)$, known as Korshunov instantons~[\onlinecite{Instantons1},\onlinecite{Instantons11}]:
\begin{equation}
e^{i\phi_{W}(\tau)} =  \sum \limits_{a=1}^{|W|}\frac{u-z_a}{1-u \bar{z}_a} , \label{eq:InstSolG}
\end{equation}
where $u = \exp(i 2\pi T\tau)$. An integer $W$ corresponds to the winding number:
\begin{equation}
W = \frac{1}{2\pi} \int\limits_{0}^{\beta}d\tau
\dot{\phi}_W .
\end{equation}
The set of complex parameters $\{z_a\}$ parametrizing the Korshunov instanton 
lie inside (outside) the unite circle, $|z_a|<1$ ($|z_a|>1$), for $W>0$ ($W<0$).
In the case of the instanton with $W=\pm 1$, one
can identify $\textrm{arg}\, z_1 /(2\pi T)$ as the position of the voltage fluctuation $i\dot{\phi}_{1}(\tau)$ in the imaginary time whereas $(1-|z_1|^2)/T$ as its duration.

The action on the Koshunov instantons is given as follows
\begin{equation}
\mathcal{S}[\phi_W]=  \frac{g}{2} |W| - i 2\pi q W +\frac{\pi^2 T}{E_c} \sum_{a,b=1}^{|W|}
\frac{1+z_a \bar{z}_b}{1-z_a \bar{z}_b}.
\label{eq:Sclass1}
\end{equation}
It is finite but explicitly depends on the set $\{z_a\}$ due to the presence of the term with the charging energy. Therefore, in the limit $E_c\gg \pi^2 T$ the configurations with  $|z_a|\to 1$ are suppressed. One could omit the last term in Eq. \eqref{eq:Sclass1} and treat the set $\{z_a\}$ as the instanton zero modes. However, it is more convenient to keep this term in the action since with this term all fluctuations around the Korshunov instanton are massive. 

As usual, the partition function can be written as a sum over different topological sectors
\begin{gather}
Z = \sum_{W=-\infty}^{\infty} Z_W,\qquad Z_W = \int_W \mathcal{D}[\phi]\, e^{-\mathcal{S}[\phi]} .\label{eq:ZSeriesW}
\end{gather}
The subscript $W$ on the integral sign denotes that the functional
integral is taken over fields with the constraint
\begin{equation}
\phi(\beta)=\phi(0)+2\pi W .\label{eq:BC}
\end{equation}
Provided relations $\bar{Z}_W =Z_{-W}$ and $Z_W \propto \exp(
2\pi i q W)$ are hold, the partition function is real and even function
of the external charge $q$. 

If one properly defines $Z_x$ as a continuation of $Z_W$ from integers to the real axis,  
then the physical observables $\mathcal{G}$ and $\mathcal{Q}$ can be written as follows
[\onlinecite{Burmistrov2}]
\begin{equation}
\begin{split}
\frac{\mathcal{G}}{4\pi} & = \im \frac{1}{2\pi i Z}
\sum_{W=-\infty}^\infty \frac{\partial Z_x}{\partial x}\Biggr
|_{x=W}, \\
\mathcal{Q} & = \re \frac{1}{2\pi i Z} \sum_{W=-\infty}^\infty
\frac{\partial Z_x}{\partial x}\Biggr |_{x=W} .
\end{split}
\label{eq:RP}
\end{equation}
In view of Eq. \eqref{eq:BC}, the relations \eqref{eq:RP} indicate that $\mathcal{G}$ and $\mathcal{Q}$ describe the response to a change in the boundary conditions. 

Restricting consideration to the gaussian fluctuations around the Korshunov instantons, one can write the partition function in a given topological sector as [\onlinecite{Instantons2,Instantons3,FKLS}]
\begin{gather}
\frac{Z_W}{Z_0} =  \left ( \frac{g^2 E_c}{2\pi^2 T}\right )^{|W|}  e^{-g |W|/2+2\pi i q W}\left (\prod\limits_{a=1}^{|W|}\int \frac{d^2 z_a}{\pi} \right )  \notag \\
\times \mathcal{J}_W(\bm{z}) \exp \left (-\frac{\pi^2 T}{E_c} \sum_{a,b=1}^{|W|}
\frac{1+z_a \bar{z}_b}{1-z_a \bar{z}_b}\right ) .
\label{eq:ZWG}
\end{gather}
Here $\bm{z}=\{z_1,\dots,z_{|W|}\}$ and the Jacobian $\mathcal{J}_W(\bm{z})$ is as follows 
\begin{equation}
\mathcal{J}_W(\bm{z}) = \frac{1}{|W|!} \det  \left |\left | \frac{1}{1-z_a \bar{z}_b} \right |\right |  .
\label{eq:JW}
\end{equation}
We note that the factor $|W|!$ takes into account that all instanton parameters $z_1, \dots, z_W$ are equivalent.

Expansion similar to Eq. \eqref{eq:ZSeriesW} can be written for a correlation function of an arbitrary operator $O$:
\begin{gather}
\langle O \rangle  = \frac{1}{Z}\sum_{W=-\infty}^{\infty} O_W,\quad O_W = \int_W \mathcal{D}[\phi]\, O[\phi] e^{-\mathcal{S}[\phi]} .\label{eq:OSeriesW}
\end{gather} 
For our purposes it will be enough to take into account the gaussian fluctuations around the Korshunov instanton in the action only. Then the quantity $O_W$ can be written similarly to Eq. \eqref{eq:ZWG}:
\begin{gather}
\frac{O_W}{Z_0} =  \left ( \frac{g^2 E_c}{2\pi^2 T}\right )^{|W|}  e^{-g |W|/2+2\pi i q W}\left (\prod\limits_{a=1}^{|W|}\int \frac{d^2 z_a}{\pi} \right )  \notag \\
\times \mathcal{J}_W(\bm{z}) O[\phi_W] \exp \left (-\frac{\pi^2 T}{E_c} \sum_{a,b=1}^{|W|}
\frac{1+z_a \bar{z}_b}{1-z_a \bar{z}_b}\right ) .
\label{eq:OWG}
\end{gather}
To the second order in  $\exp(-g/2)$, we can write
\begin{gather}
\langle O \rangle = \langle O \rangle^{(0)} + \langle O \rangle^{(1)}+ \langle O \rangle^{(2)}+\dots
\label{eq:OO}
\end{gather}
where $\langle O \rangle^{(0)} = O_0/Z_0$, 
\begin{equation}
\langle O \rangle^{(1)} = \bigl (O_1+O_{-1}-\langle O \rangle^{(0)} (Z_1+Z_{-1})\bigr )/Z_0  ,
\end{equation} 
and
\begin{gather}
\langle O \rangle^{(2)} =  \frac{O_2+O_{-2}-\langle O \rangle^{(0)} (Z_2+Z_{-2})}{Z_0} \notag \\
- \langle O \rangle^{(1)}\,  \frac{Z_1+Z_{-1}}{Z_0} .
\label{eq:OO2}
\end{gather}
We are interested in the two-point correlation function \eqref{eq:Ddef}. Evaluation 
of $\langle O \rangle^{(1)}$ for $D(i\omega_n)$ was performed in Ref. [\onlinecite{Burmistrov2}].  Below we remind this computation first, and then evaluate $\langle O \rangle^{(2)}$.

\subsection{One-instanton contribution $\langle O \rangle^{(1)}$}

Using Eqs. \eqref{eq:ZWG} and \eqref{eq:JW},  we can write $Z_1/Z_0$ as 
\begin{equation}
\frac{Z_1}{Z_0}  = \frac{g^2 E_c}{2\pi^2 T} e^{-g /2+2\pi i q} \int\limits_{|z|<1} \frac{d^2 z}{\pi} \frac{1}{1-|z|^2}
e^{-\frac{\pi^2 T}{E_c} \frac{1+|z|^2}{1-|z|^2}} .
\end{equation}
Evaluating the integral over $z$ in the limit $E_c\gg \pi^2 T$, we find
\begin{equation}
\frac{Z_1}{Z_0}  =  \frac{g^2 E_c}{2\pi^2 T} e^{-g /2+2\pi i q} \ln \left (\frac{E_c e^{-\gamma}}{2\pi^2 T}\right )
, \label{eq:Z1p}
\end{equation}
where $\gamma \approx 0.577$ denotes the Euler constant. The classical value of the two-point correlation function of the Coulomb boson is given as
\begin{equation}
D(i\omega_n|\phi_{\pm 1}) = \beta \oint
 \frac{d u du^\prime}{(2\pi i)^2 u u^\prime} u^n {u^\prime}^{-n}
\frac{u-z}{1-u \bar{z}} \frac{1-u^\prime \bar{z}}{u^\prime-z} .
\end{equation}
Here the integrals are assumed over the unit circles: $|u|=1$ and $|u^\prime|=1$. Performing integration over $u$ and $u^\prime$, one finds
\begin{equation}
D(i\omega_n|\phi_{\pm 1}) =  \beta \Bigl [ |z|^2 \delta_{n,0} + (1-|z|^2)^2 |z|^{2(|n|-1)} \theta(\pm n)\Bigr ] ,
\end{equation}
where $\theta(x)$ denotes the Heaviside step function with $\theta(0)=0$. We note that the classical value of $D(i\omega_n)$ on the solution $\phi_1$ ($\phi_{-1}$) vanishes for $\omega_n<0$ ($\omega_n>0$).
Next evaluating integral over $z$ in Eq. \eqref{eq:OWG} we obtain
\begin{gather}
\langle D(i\omega_n)\rangle^{(1)} = - \beta \frac{g^2 E_c}{2\pi^2 T} e^{-g /2}  \Biggl [ 2 \cos(2\pi q) \delta_{n,0}
\notag \\
-  e^{2\pi i q\sgn \omega_n}\frac{(2\pi T)^2 (1-\delta_{n,0})}{|\omega_n|(|\omega_n|+2\pi T)}\Biggr ] .
\label{Eq:D1i}
\end{gather}
Performing summation over Matsubara frequencies in Eq. \eqref{eq:Kdef}, from Eq. \eqref{Eq:D1i} we find the following one-instanton correction to the response function $K(i\omega_n)$ [\onlinecite{Burmistrov2}]:
\begin{gather}
\langle K(i\omega_n)\rangle^{(1)} =
 \frac{g^3 E_c}{2\pi^2} e^{-g/2} \Biggl \{ e^{-2\pi i q} \Bigl [  \psi\Bigl (1+\frac{\omega_n}{2\pi T}\Bigr )-\psi(1)\Bigr ] \notag \\
  - \cos(2\pi q) \sum\limits_{m=2}^\infty \frac{1}{m} \Biggr \}  .
\label{eq:K1i}  
\end{gather}
Here $\psi(z)$ stands for the digamma function.

\subsection{Two-instanton contribution}

Using Eqs. \eqref{eq:ZWG} and \eqref{eq:JW},  we can write $Z_2/Z_0$ as 
\begin{align}
\frac{Z_2}{Z_0}  = & \frac{1}{2} \left (\frac{g^2 E_c}{2\pi^2 T} \right )^2 e^{-g +4\pi i q} \int\limits_{|z_{1,2}|<1} \frac{d^2 z_1 d^2z_2}{\pi^2} F_-(z_1,z_2)  \notag \\
& \times 
\exp\Bigl [-\frac{2\pi^2 T}{E_c} (1-|z_1z_2|^2) F_+(z_1 z_2) \Bigr] .
\end{align}
Here we introduce two functions
\begin{equation}
F_\pm (z_1,z_2)= \frac{1}{(1-|z_1|^2)(1-|z_2|^2)}\pm \frac{1}{(1-z_1\bar{z}_2)(1-\bar{z}_1z_2)} .
\end{equation}
Evaluating the integrals over $z_1$ and $z_2$ under the following assumption, $E_c\gg \pi^2 T$, we find
\begin{equation}
\frac{Z_2}{Z_0}  = \frac{1}{2} \left (\frac{g^2 E_c}{2\pi^2 T}\right )^2 e^{-g +4\pi i q} \Bigl [ \ln^2 \left (\frac{E_c e^{-\gamma}}{2\pi^2 T}\right ) -\frac{\pi^2}{6}\Bigr ]
. \label{eq:Z2p}
\end{equation}
The classical value of the two-point correlation function of the Coulomb boson on the two-instanton solution $\phi_{\pm 2}$ is given as
\begin{gather}
D(i\omega_n|\phi_{\pm 2}) = \beta \oint
 \frac{d u du^\prime}{(2\pi i)^2 u u^\prime} u^n {u^\prime}^{-n}
\frac{u-z_1}{1-u \bar{z}_1}\frac{u-z_2}{1-u \bar{z}_2} \notag \\
\times \frac{1-u^\prime \bar{z}_1}{u^\prime-z_1} 
 \frac{1-u^\prime \bar{z}_2}{u^\prime-z_2}  .
\end{gather}
As above, the integrals are assumed over the unit circles: $|u|=1$ and $|u^\prime|=1$. Performing integration over $u$ and $u^\prime$, one finds
\begin{gather}
D(i\omega_n|\phi_{\pm 2}) =  \beta \Biggl [ \frac{\theta(\pm n)}{|z_1-z_2|^2}
\Biggl | z_1^{|n|-1} (1-|z_1|^2) (1-z_1\bar{z}_2) 
\notag \\
 -  z_2^{|n|-1} (1-|z_2|^2) (1-\bar{z}_1 z_2) \Biggr |^2+|z_1z_2|^2 \delta_{n,0}  \Biggr ] .
\end{gather}
We note that the classical value of $D(i\omega_n)$ on the solution $\phi_2$ ($\phi_{-2}$) vanishes for $\omega_n<0$ ($\omega_n>0$). Next, evaluating the integrals over $z_1$ and $z_2$ in Eq. \eqref{eq:OWG}, we obtain
\begin{gather}
\frac{D_2-\langle D\rangle^{(0)} Z_2}{Z_0} =\frac{g^4 E_c^2}{4\pi^4 T} e^{-g+4\pi i q}  \Biggl \{ \ln\left (\frac{E_c e^{-1-\gamma}}{2\pi^2 T}\right )  \delta_{n,0} 
\notag \\
-  \theta(n) \Bigl [\frac{1}{n(n+1)}\ln\left (\frac{E_c e^{-\gamma}}{2\pi^2 T}\right )  - \frac{2n+1}{n^2(n+1)^2} \Bigr ]\Biggr \} .
\end{gather}
Hence, using Eq. \eqref{eq:OO2}, we find
\begin{gather}
\langle D(i\omega_n)\rangle^{(2)} =  \frac{g^4 E_c^2}{4\pi^4 T} e^{-g }  \Biggl \{  \Bigl [ 2\delta_{n,0} -  \frac{(2\pi T)^2(1-\delta_{n,0})}{|\omega_n|(|\omega_n|+2\pi T)}\Bigr ] \notag\\
\times \ln\left (\frac{E_c e^{-\gamma}}{2\pi^2 T}\right ) 
-   (1-\delta_{n,0})  e^{4\pi i q\sgn \omega_n}(2\pi T)^2 \Bigl [ \frac{1}{\omega_n^2} \notag \\
- \frac{1}{(|\omega_n|+2\pi T)^2}\Bigr ]+2 \delta_{n,0} \cos(4\pi q)\Biggr \} .
\label{Eq:D2i}
\end{gather}

Performing summation over Matsubara frequencies in Eq. \eqref{eq:Kdef}, we find the following two-instanton correction to the response function $K(i\omega_n)$:
\begin{gather}
\langle K(i\omega_n)\rangle^{(2)} =
 \frac{g^5 E_c^2}{4\pi^4 T} e^{-g} \Biggl \{ e^{-4\pi i q} \Bigl [  \psi^\prime\Bigl (1+\frac{\omega_n}{2\pi T}\Bigr )-\psi^\prime(1)\Bigr ] \notag \\
  + \frac{\pi^2}{6} \cos (4\pi q)  - \ln\left (\frac{E_c e^{-\gamma}}{2\pi^2 T}\right )  \Bigl [\psi\Bigl (1+\frac{\omega_n}{2\pi T}\Bigr ) \notag \\
  -\psi (1)
- \sum\limits_{m=2}^\infty \frac{1}{m} \Bigr ]\Biggr \}  .
\label{eq:K2i}  
\end{gather}

It is worthwhile to mention that the two-instanton correction \eqref{eq:K2i} to the response function $K(i\omega_n)$ contains terms which are independent of $q$. As we demonstrate in Appendix \ref{app1}, these terms are cancelled by the contribution to $\langle K(i\omega_n)\rangle^{(0)}$ due to configurations with one instanton and one anti-instanton. Although, this configuration is not the exact solution of the classical equation of motions for $\mathcal{S}_d$, it provides a significant contribution to $\langle K(i\omega_n)\rangle^{(0)}$. Also, we mention that such $q$ independent terms do not contribute to the Coulomb blockade oscillations. Therefore, we neglect them in what follows.

\subsection{Final results}

Using Eqs. \eqref{eq:Z1p} and \eqref{eq:Z2p}, the partition function can be written as 
\begin{gather}
\ln Z = \ln Z_0 + \frac{g^2 E_c}{\pi^2 T} e^{-g/2} \ln\left (\frac{E_c e^{-\gamma}}{2\pi^2 T}\right ) \cos(2\pi q) \notag \\
- \frac{g^4 E_c^2}{24\pi^2 T^2} e^{-g} \Biggl [ \cos(4\pi q) + \frac{6}{\pi^2}  \ln^2\left (\frac{E_c e^{-\gamma}}{2\pi^2 T}\right ) \Biggr ]  .
\end{gather}
We note that this result coincides with the result of Ref. [\onlinecite{FKLS}].  With the help of Eq. \eqref{eq:def-avQ}, we find the average charge on the island within two-instanton approximation:
\begin{align}
Q = & q - \frac{g^2}{\pi} e^{-g/2} \ln\left (\frac{E_c e^{-\gamma}}{2\pi^2 T}\right ) \sin(2\pi q)\notag \\
& + \frac{g^4 E_c}{12\pi^2 T} e^{-g} \sin(4\pi q) .
\label{eq:final-avQ}
\end{align}

Combining together Eqs. \eqref{eq:K1i} and \eqref{eq:K2i}, and performing analytic continuation to the real frequencies, $i \omega_n \to \omega+i0^+$, we  find 
\begin{align}
K^R(\omega) &=  \frac{i\omega g}{4\pi} \left [ 1- \frac{2}{g} \ln\left (\frac{e g E_c}{2\pi^2 T}\right ) 
 +\frac{2}{g} \psi \left (1 - \frac{i\omega}{2\pi T}\right ) \right ] \notag \\
  &+ \frac{g^3 E_c}{2\pi^2} e^{-g/2 -2\pi i q} \left [\psi \left (1 - \frac{i\omega}{2\pi T}\right ) - \psi(1) \right ] 
\notag \\
 &+ \frac{g^5 E_c^2}{4\pi^4 T} e^{-g-4\pi i q} \left [\psi^\prime \left (1 - \frac{i\omega}{2\pi T}\right )-\psi^\prime(1)  \right ] .
\label{eq:finalK}
\end{align}
Here we add the perturbative result (the first line in Eq. \eqref{eq:finalK}) [\onlinecite{perturb}]. Using Eq. \eqref{eq:defG} we obtain the conductance of the SET within the two-instanton approximation:
\begin{gather}
\mathcal{G} = g  \left [ 1- \frac{2}{g} \ln\left (\frac{g E_ce^{1+\gamma}}{2\pi^2 T}\right ) \right ] - \frac{g^3 E_c}{6 T} e^{-g/2} \cos(2\pi q) \notag \\
 +\frac{\zeta(3) g^5 E_c^2}{\pi^4 T^2} e^{-g} \cos(4\pi q)  .
\label{eq:finalG}
\end{gather}
Here $\zeta(z)$ stands for the Riemann zeta function. As one can see from Eq. \eqref{eq:finalG} the expansion over topological sectors in the conductance is controlled by the parameter $(g^2 E_c/T) \exp(-g/2) \ll 1$. In addition, we note small numerical factors appearing in one- and two-instanton contributions: $1/6$ and $\zeta(3)/\pi^4 \approx 0.01$, respectively.   

Using Eq. \eqref{eq:defQ}, we obtain the effective charge of the SET within the two-instanton approximation:
\begin{align}
\mathcal{Q} = & q - \frac{g^3 E_c}{24\pi T} \Bigl [ 1+\frac{24 T}{E_c}\ln\left (\frac{E_c e^{-\gamma}}{2\pi^2 T}\right ) \Bigr ]e^{-g/2} \cos(2\pi q) \notag  \\
& +\frac{\zeta(3) g^5 E_c^2}{4\pi^5 T^2} \Bigl [ 1+ \frac{\pi^4 T}{3\zeta(3)E_c}\Bigr ]e^{-g} \cos(4\pi q)  .
\label{eq:finalQ0}
\end{align}
We note that corrections to unity in the square brackets in the right hand side of Eq. \eqref{eq:finalQ0} appear due to the presence of the average charge in Eq. \eqref{eq:defQ}. Within our assumption, $E_c\gg \pi^2 T$, we can safely neglect them and obtain
\begin{gather}
\mathcal{Q} =  q - \frac{g^3 E_c}{24\pi T}  e^{-g/2} \sin(2\pi q)  +\frac{\zeta(3) g^5 E_c^2}{4\pi^5 T^2} e^{-g} \sin(4\pi q)  .
\label{eq:finalQ}
\end{gather}

\section{Discussion and conclusions}
\label{s4}

The result  \eqref{eq:finalK} for the response function allows us to determine the current-voltage characteristic of a SET within the two-instanton approximation. Introducing two currents 
\begin{equation}
I_{s,d} = -\frac{2 g_{s,d}}{g_s+g_d}\im K^R(-V_{s,d}) ,
\label{eq:Isd}
\end{equation}
one can find dependence of the current $I$ through a SET on the bias voltage $V= V_s-V_d$ from the current conservation conditions 
\begin{equation}
I = I_s =  - I_d .
\label{eq:CVC0}
\end{equation}
We mention that for evaluation of the right hand side of Eq. \eqref{eq:Isd} with the help of Eq. \eqref{eq:finalK} one needs to make the following substitution: $q \to q(V) = C_g V_g + (C_s V_s + C_d V_d)$. We emphasize that the dependence of the external charge on the bias voltage becomes important at $|V| \gtrsim E_c$ and makes the oscillations to be more of non-sinusoidal type.

In the special case of asymmetric SET, $g_s\ll g_d$, we find the following result for the current at low bias voltages, $|V|\ll E_c$:
\begin{equation}
I(V) = -2 \frac{g_s}{g} \im K^R(-V)
\label{eq:IV1}
\end{equation}
where $K^R(V)$ is given by Eq. \eqref{eq:finalK}. Here we assume also that $C_{s,d} \sim C$.
As one can see from Eq. \eqref{eq:finalK} the amplitudes of harmonics of the Coulomb blockade oscillations are controlled by a small parameter $g^2 E_c e^{-g/2}/\max\{2\pi T,|V|\} \ll 1$. Using Eq. \eqref{eq:IV1}, we obtain the following explicit expression for the differential conductance $\partial I/\partial V = (e^2/h) (g_s/g)\mathcal{G}(V)$ at $2\pi T\ll |V| \ll E_c$:
\begin{gather}
\mathcal{G}(V)=g \Biggl [ 1- \frac{2}{g} \ln\left (\frac{g E_c}{\pi V}\right ) 
+\frac{2g^2 E_c}{\pi V} e^{-g/2} \sin(2\pi q) \notag \\
+ \frac{2g^2 E_c T}{V^2} e^{-g/2} \cos(2\pi q)- \frac{2g^4 E^2_c}{\pi^2 V^2} e^{-g} \cos(4\pi q) \Biggr ] .
\label{eq:DifCond}
\end{gather}
This result implies that the increase of the bias voltage $V$ beyond $2\pi T$ results in suppression of the Coulomb blockade oscillations. In addition the bias voltage leads to a $V$ dependent shift of minima and maxima 
of oscillations. 

In recent experiments  [\onlinecite{Frydman2011},\onlinecite{Frydman2015}] the electron transport via highly asymmetric SET at low temperatures was studied. For several samples with $1 < g < 10$ distinct Coulomb blockade oscillations of the conductance have been measured. The following features have been observed: (i) higher harmonics of oscillations of $G$ become visible with decrease of the total tunneling conductance $g$; (ii) logarithm of the amplitude of the first harmonic is linear function of $g$; (iii) amplitudes of harmonics are not monotonously suppressed with their number in a sample with $g \sim 3$; (iv) the bias voltage $V$ suppresses higher harmonics of the Coulomb blockade oscillations; (v) no visible $V$-dependent phase shift of oscillations were reported. We note that although items (i), (ii) and (iv) are in qualitative agreement with our theory, observations (iii) and (v) are formally at odds with the observations of Refs. [\onlinecite{Frydman2011},\onlinecite{Frydman2015}]. We note that feature (iii) can be related with the fact that the total tunneling conductance is not large enough ($g\sim 3$) such that application of our theory is questionable.

As future perspectives, it would be interesting to extend our theory to the non-equilibrium conditions, e.g. to take into account that the electron distribution function on the island is different from the Fermi distribution due to the presence of a bias voltage. Perhaps, this could be done with the help of recent non-equilibrium generalization of the Koshunov instantons [\onlinecite{GutmanTitov}] and the kinetic equation for the AES model [\onlinecite{RBC}].

In summary, we compute the response function within the two-instanton approximation to the AES model. This allows us to determine the temperature and gate voltage dependence of the Coulomb blockade oscillations in the conductance and the effective charge. In agreement with Ref. [\onlinecite{Cond}], we found that the amplitudes of harmonics of the Coulomb blockade oscillations of the conductance and the effective charge is controlled by a small parameter, $[g^2 E_c e^{-g/2}/(2\pi T)] \ll 1$. In particular, this implies that the amplitude of the second harmonic is much smaller than the amplitude of the first one. For a small bias voltage, $|eV| \ll E_c$, the amplitude of the Coulomb blockade oscillations in the differential conductance is controlled by a small parameter  $g^2 E_c e^{-g/2}/\max\{2\pi T, |eV|\} \ll 1$. A finite bias voltage leads to suppression of the Coulomb blockade oscillations and the appearance a $V$ dependent phase shift. Our results allows to qualitatively understand some features of experimental findings of Refs. [\onlinecite{Frydman2011},\onlinecite{Frydman2015}].

\begin{acknowledgments}
We acknowledge useful discussions with A. Frydman, D. Gutman, and Ya. Rodionov. The research was funded by Russian Science Foundation under the grant No. 14-12-00898.
\end{acknowledgments}

\vspace{1cm}
\appendix

\section{Configuration with one instanton and one anti-instanton}
\label{app1}

In this appendix we consider the contribution to the correlation function $D(i\omega_n)$ due to configuration with one instanton and anti-instanton: 
\begin{equation}
e^{i\phi_{1,-1}} = \frac{u - {z}_1}{1-u \bar{z}_1}\frac{1-u \bar{z}_2}{u - z_2} ,\qquad |z_{1,2}|\leqslant 1 .
\label{eq:App:1-1}
\end{equation} 
We note that such configuration is not the solution of the classical equation of motion for the action \eqref{eq: SdStart}.The classical action is given as
\begin{gather}
\mathcal{S}[\phi_{1,-1}] = g \Biggl ( 1 - \frac{(1-|z_1|^2)(1-|z_2|^2)}{|1-z_1\bar{z}_2|^2}\Biggr ) \notag \\
+ \frac{2\pi^2 T}{E_c} F_-(z_1,z_2) .
\end{gather}
The classical value of the two-point correlation function of the Coulomb boson is given as
\begin{gather}
D(i\omega_n|i\phi_{1,-1})= \beta \oint
 \frac{d u du^\prime}{(2\pi i)^2 u u^\prime} u^n {u^\prime}^{-n}
\frac{u-z_1}{1-u \bar{z}_1}\frac{1-u \bar{z}_2}{u-z_2} \notag \\
\times \frac{1-u^\prime \bar{z}_1}{u^\prime-z_1} 
 \frac{u^\prime-z_2}{1-u^\prime \bar{z}_2}  .
\end{gather}
Performing integrations over $u$ and $u^\prime$, we find
\begin{align}
D(i\omega_n|\phi_{1,-1}) & =\beta \Biggl [\delta_{n,0} \Biggl |\frac{1-|z_1|^2-|z_2|^2+z_1 \bar{z}_2}{1-z_1 \bar{z}_2}\Biggr |^2
\notag \\
+ & \theta(-n) |z_2|^{2(|n|-1)} (1-|z_2|^2)^{2}\Biggl | \frac{z_1-z_2}{1-z_1\bar{z}_2}\Biggr |^2
\notag \\
+ & \theta(n) |z_1|^{2(n-1)} (1-|z_1|^2)^{2}\Biggl | \frac{z_1-z_2}{1-z_1\bar{z}_2}\Biggr |^2  \Biggr ] .
\end{align}
After evaluation of integrals over $z_1$ and $z_2$ with logarithmic accuracy, we find the contribution due to configuration of instanton and anti-instanton:
\begin{align}
\langle D(i\omega_n)\rangle^{(1,-1)} & = \frac{g^4 E_c^2}{4\pi^4 T} e^{-g}\Biggl \{ 
-2 \delta_{n,0} + (2\pi T)^2 ( 1-\delta_{n,0}) \notag \\
& \times \frac{1}{|\omega_n|(|\omega_n|+2\pi T)} \Biggr \} \ln \frac{E_c e^{-\gamma}}{2\pi^2 T} .
\end{align}
We note that in this case the factor $1/|W|!$ in the Jacobian \eqref{eq:JW} is absent since parameters $z_1$ and $z_2$ in the ansatz \eqref{eq:App:1-1} are distinguishable. As one can see, configuration with instanton and anti-nstanton provides the contribution to the two-point correlation function which cancels exactly the term proportional to $\ln [{E_c e^{-\gamma}}/{(2\pi^2 T)}]$ in the contribution due to 2 instantons or 2 anti-instanons, cf. Eq. \eqref{Eq:D2i}.


\end{document}